Vestoids, Part II: The basaltic nature and HED meteorite analogs for eight $V_p$-type asteroids and their associations with (4) Vesta

Paul S. Hardersen[1,*], Vishnu Reddy[2], Rachel Roberts[3]

[1] University of North Dakota, Department of Space Studies, 4149 University Avenue, Stop 9008, 530 Clifford Hall, Grand Forks, ND 58202-9008, USA, Hardersen@space.edu

[*] Visiting Astronomer at the Infrared Telescope Facility, which is operated by the University of Hawai'i under contract NNH14CK55B with the National Aeronautics and Space Administration

[2] Planetary Science Institute, 1700 E. Fort Lowell Road, Suite 106, Tucson, AZ 85719, USA

[3] University of North Dakota, Department of Space Studies, 4149 University Avenue, Stop 9008, 521 Clifford Hall, Grand Forks, ND 58202-9008, USA

# Characterization of eight $V_p$-type asteroids

Abstract. Improving constraints on the abundance of basaltic asteroids in the main asteroid belt is necessary for better understanding the thermal and collisional environment in the early solar system, for more rigorously identifying the genetic family for (4) Vesta, for determining the effectiveness of Yarkovsky/YORP in dispersing asteroid families, and for better quantifying the population of basaltic asteroids in the outer main belt ($a > 2.5$ AU) that are likely unrelated to (4) Vesta.

NIR spectral observations in this work were obtained for the $V_p$-type asteroids (2011) Veteraniya, (5875) Kuga, (8149) Ruff, (9147) Kourakuen, (9553) Colas, (15237) 1988 $RL_6$, (31414) Rotaryusa, and (32940) 1995 $UW_4$ during August/September 2014 utilizing the SpeX spectrograph at the NASA Infrared Telescope Facility (IRTF), Mauna Kea, Hawai'i. Spectral band parameter (band centers, Band Area Ratios) and mineralogical analysis (pyroxene chemistry) for each average asteroid NIR reflectance spectrum suggests a howardite-eucrite-diogenite (HED) meteorite analog for each asteroid. (5875) Kuga is most closely associated with the eucrite meteorites, (31414) Rotaryusa is most closely associated with the diogenites, and the remaining other six asteroids are most closely associated with the howardite meteorites. Along with orbital locations in the inner main belt and in the vicinity of (4) Vesta, the existing evidence suggests that these eight $V_p$-type asteroids are also likely Vestoids.

1. Introduction.

The scientific impact of the visit of the Dawn spacecraft to main-belt asteroid (4) Vesta continues to flow as analysis of Dawn data continues to produce surprises about the asteroid. Recent analysis has shown evidence for the presence of likely exogenic meteoritic impact debris on Vesta with both carbonaceous and olivine-bearing origins (Reddy et al., 2012a; Le Corre et al., 2015; Nathues et al., 2015; Palomba et al., 2015; Poulet et al., 2015). Despite reconfirmation that Vesta's surface is predominantly howarditic of a generally basaltic origin (De Sanctis et al., 2012), the lack of expected upper mantle olivine exposures in the large south pole impact region is causing reconsideration of petrogenetic models describing Vesta's formation (Palomba et al., 2015). Accumulating physical data that is incongruent with existing formation models has also led to questions about fundamental assumptions of Vesta's nature as a fully differentiated proto-planet (Consolmagno et al., 2015).

Vesta's collisional history has also been documented by crater counts that have identified 1872 craters ($D > 4$ km) with 12 craters having diameters greater than 50 km (Marchi et al., 2012). There should be some relationship between the cratering record on Vesta and the impact debris that has escaped the asteroid during the formation and subsequent epochs, but the passage of time and small body orbital perturbations (Yarkovsky/YORP) make a direct connection difficult (Bottke et al., 2006). Nonetheless, the development of asteroid taxonomies (i.e., V-type) and the Vesta families has provided initial constraints on a relatively large pool of asteroids that were potentially excavated from Vesta's surface (Zappala et al., 1995; Mothe'-Diniz et al., 2012; Nesvorny, 2012).

However, providing a better understanding of Vesta's collisional history, cratering record, and evolution requires a more accurate inventory of the basaltic asteroid population near Vesta. Initial work has mostly focused on the first Vestoids that were discovered in orbital locations between Vesta and the 3:1 mean motion resonance, but investigation of some other potential Vestoids has also occurred (Binzel and Xu, 1993; Vilas et al., 2000; Burbine et al., 2001; Kelley et al., 2003; Cochran et al., 2004; Alvarez-Candal et al., 2006; Duffard et al., 2004, 2006; Roig et al., 2008; Duffard and Roig, 2009; Moskovitz et al., 2010; De Sanctis et al., 2011a,b; Mayne et al., 2011; Reddy et al., 2011; Solontoi et al., 2012; Hicks et al., 2014). Hardersen et al. (2014) reported the first efforts of a relatively new, large study to more accurately identify basaltic asteroids in the main asteroid belt with a primary effort to better determine those basaltic asteroids most likely to be associated with Vesta. While asteroid taxonomic classes can be suggestive of surface compositions and meteorite analogs, they are not explicitly designed for compositional identifications and have inherent ambiguities that can lead to compositional mis-interpretations (Tholen, 1984; Carvano et al., 2010).

Through the combination of existing asteroid taxonomic classifications ($V_p$-class: Carvano et al. (2010)) and Wide-Field Infrared Survey Explorer (WISE) albedo and diameter data, Mainzer et al. (2012) identified a group of ~600 asteroids that exhibit evidence favoring a basaltic composition and originating from the surface of Vesta.

Hardersen et al. (2014) then began a compositional study of the ~600 $V_p$-type asteroids in Mainzer et al. (2012) by adding near-infrared (NIR) spectral and mineralogic evidence, coupled with meteorite analog identification, to more rigorously test and interpret eight $V_p$-type asteroids as most probably originating from the surface of (4) Vesta.

This work reports on an additional eight $V_p$-class asteroids [(2011) Veteraniya, (5875) Kuga, (8149) Ruff, (9147) Kourakuen, (9553) Colas, (15237) 1988 RL$_6$, (31414) Rotaryusa, (32940) 1995 UW$_4$] to determine their possible genetic association with, and origin from, Vesta. A basaltic asteroid that likely originates from Vesta's surface is termed a Vestoid, which was collisionally ejected by impacts to become a part of the Vesta family. Of all the taxonomic classes defined by Carvano et al. (2010), the $V_p$-class is most suggestive of a Vesta-like visible-wavelength spectrum due to the presence of the deepest putative 0.9-μm absorption feature. The $V_p$-class asteroids exhibit the highest WISE-derived albedos ($p_v = 0.343 \pm 0.010$) of any Carvano et al. (2010) taxonomic class and they are also consistent with albedos for basaltic achondrite meteorites and Vesta (Mainzer et al., 2012). The $V_p$ asteroids in this paper are also located inside the 3:1 mean-motion resonance ($a < 2.5$ AU) in the orbital neighborhood of (4) Vesta. Details on the $V_p$ taxonomic class and WISE albedo data can be found in Hardersen et al. (2014).

These existing data are supplemented with NIR spectral data (0.7 to 2.5-μm) that exhibit diagnostic absorption features in the ~0.9- and ~1.9-μm spectral regions indicative of pyroxene group minerals due to the presence of $Fe^{2+}$ in the mineral crystal structures (Burns, 1993). Pyroxene is the most important mineral group for the study of basaltic asteroids as it is the sole, major mafic silicate mineral present in basalts. The basaltic achondrites in the terrestrial meteorite record primarily consist of the howardite-eucrite-diogenite (HED) meteorite group that is thought to derive from Vesta's surface, although some anomalous and possibly non-Vesta-related basaltic achondrites have been found (Bland et al., 2009).

The HED meteorites exhibit systematic mineralogic and spectral trends from the eucrites to the howardites to the diogenites. Eucrite major minerals include inverted pigeonite and calcic plagioclase with ferrosilite (Fs) abundances generally ranging from Fs ~40-70 (Mittlefehldt, 1998). Diogenites' sole major mineral is orthopyroxene with iron abundances ranging from Fs ~23-28 (Mittlefehldt et al., 1998). Howardite major mineral proportions and chemistries are a function of their eucrite and diogenite components and exhibit intermediate spectral and mineralogic trends between the eucrites and diogenites (Mittlefehldt et al., 1998). A trend of increasing relative pyroxene:feldspar abundance is seen in the eucrite-to-diogenite sequence, which also manifests in the disappearance of a weak absorption feature in the ~1.2-μm region with steadily decreasing abundances of Fs within the pyroxenes (Klima et al., 2007). Plagioclase feldspar is the other major mineral in basalts that also displays a weak, ephemeral ~1.2-μm absorption, but its spectral contribution in HED meteorite and basaltic asteroid NIR spectra is uncertain (Hardersen et al., 2006).

2. Observations.

NIR spectra for the eight $V_p$-type asteroids reported here were obtained at the NASA Infrared Telescope Facility (IRTF), Mauna Kea, Hawai'i, on August 26, September 3, and September 4, 2014 UT. Observations utilized the upgraded SpeX 0.7-5.3-μm medium resolution spectrograph and imager (http://irtfweb.ifa.hawaii.edu/~spex). The primary upgrades completed in 2014 include replacing the infrared arrays and much of the warm electronics hardware. Details on the original SpeX instrument and performance parameters are found in Rayner et al. (2003, 2004). These asteroids were chosen based on observational constraints such as availability during the assigned observing time, brightness, and position in the sky at relatively low airmasses.

NIR spectra were obtained using the SpeX prism mode across the wavelength range from ~0.7-2.5-μm. Spectra were obtained at a spectral resolution, $R = \lambda/\Delta\lambda \sim 95$. All observations were obtained at the parallactic angle utilizing the 0.8" slit. Flat field images and argon arc spectra calibration frames were obtained using the internal SpeX mechanical setup and obtained via a macro computer program specifically designed for the chosen SpeX observing configuration.

Our observational goal is to obtain at least 10 NIR spectra per $V_p$-type asteroid to ensure a sufficient signal-to-noise-ratio (SNR) in each reduced average asteroid NIR reflectance spectrum and to allow the observation of ~8-10 $V_p$-type asteroids per full night. Each $V_p$-type asteroid is paired with a non-variable G-type main sequence star that is usually within ~3° of the asteroid during observations. This close proximity increases the likelihood that atmospheric conditions during asteroid and stellar observations are similar, which improves the ability to later correct for telluric absorptions in the asteroid NIR spectra. The stellar (i.e., extinction star) observations are sequenced to bracket the asteroid observations with the stellar spectra being acquired over a finite, but larger, airmass range than the airmass range during the asteroid observations. Observations of individual objects are acquired using an alternating sequence where the slit position alternates between the $A$- and $B$-beam positions in an $ABBAABBA...$ sequence. The low resolution spectral observations ($R \sim 95$) with the 0.8" slit, and at the parallactic angle, also facilitates relatively consistent guiding that minimizes spectral slope variations. This overall sequence allows efficient background sky subtraction by pairing asteroid and stellar spectra during the data reduction process.

Ten NIR spectra were acquired for seven of the $V_p$-type asteroids while 14 spectra were obtained for (32940) 1995 UW$_4$. Two sets of observations were made of (15237) 1988 RL$_6$ on different nights (September 3 and 4, 2014 UT) with 10 NIR spectra being obtained each night. In addition, 10-30 NIR spectra were obtained nightly for the solar analog star, HD 28099. Solar analog star observations are required to correct overall asteroid NIR spectral slopes when non-G2V extinction stars are used with an asteroid. Table 1 summarizes the observational details of the $V_p$-type asteroids, extinction stars, and solar analog star in this work. Table 2 provides basic ephemeris and physical information for the eight $V_p$-type asteroids. Figure 1 presents an inclination vs. semi-major axis plot for the $V_p$-type asteroids in this work, along with plots for (4) Vesta and the overall ~600 $V_p$-type asteroid dataset for this project.

3. Data Reduction and Analysis.

Data reduction to produce an average NIR reflectance spectrum for each asteroid involves the use of Spextool and Microsoft Excel (Cushing et al., 2004; http://irtfweb.ifa.hawaii.edu/~spex/). Spextool is an IDL-based data reduction software package specifically designed for SpeX whose primary features include background sky removal, wavelength calibrations, flat fielding, telluric corrections, sub-channel pixel shifting, averaging, and display functions. Spextool is multi-functional and allows manual user control in every step of the process to ensure all of the ingested data can be individually examined and, if necessary, excluded from the data reduction process.

Wavelength calibration requires at least one argon arc spectrum that converts CCD channels to wavelengths. Calibration, which is performed automatically, occurs by identifying discrete argon emission lines on the array and matching emission line locations on the array with known argon emission line wavelengths. At least five flat field images are also acquired and averaged to produce an average flat field image that removes pixel-to-pixel sensitivity variations in the array. These calibrations are applied to all acquired spectra.

Background subtraction requires the pairing of spectra acquired during the observing run. This allows the efficient removal of the background sky signal as the sky signal should be near constant in the few second to minutes necessary to acquire two NIR spectra of a star or asteroid. On a given observing night, two sets of stellar spectra are acquired with 10 spectra per set, 10-30 solar analog spectra are acquired, and at least 10 spectra per asteroid are acquired.

Telluric corrections utilize the two sets of extinction star observations that were acquired before and after the associated asteroid spectra. The two sets of stellar spectra are averaged, which are then ratioed to each individual asteroid spectrum (i.e., asteroid/avg. stellar spectrum) to correct for the primary atmospheric $H_2O$ vapor absorptions at ~1.4- and ~1.9-μm. Optimum telluric corrections typically occur when spectra are obtained on photometric nights with a stable atmosphere (i.e., clear skies, good seeing) and spectra are acquired at low airmass. Telluric correction for the ~1.4-μm absorption is usually more effective than the ~1.9-μm absorption. The effectiveness of the applied corrections can be qualitatively assessed by comparing the point-to-point scatter in both regions of the asteroid spectrum. The telluric correction process also provides the ability to automatically or manually offset the stellar average spectrum and each asteroid spectrum to correct for minor array shifting due to telescope flexure and movement.

The same data reduction process above is applied to the solar analog star spectra. Average spectra for each asteroid, and for the solar analog star, are then created and analyzed for quality across the entire spectral range and in the regions of the telluric absorptions. Low quality spectra (i.e., significant point-to-point data scatter) or spectra that show abnormalities are omitted from the averaging process.

All average asteroid and solar analog spectra are then converted to text files and imported into Microsoft Excel, where the primary tasks involve smoothing the average solar analog spectrum, performing the final average asteroid-to-solar analog spectrum ratio, and normalizing each final asteroid spectrum to 1.5-μm. Smoothing of the solar analog spectrum is necessary due to the common presence of residual telluric absorption features and the expectation that broad NIR stellar absorption features are not present in the ~0.7-to-2.5-μm spectral region.

Construction of the final average spectrum for each asteroid can be summarized in the following equation:

$$Avg.\ Asteroid/Solar\ Analog = (Asteroid/Ext.\ Star)\ /\ (Solar\ Analog/Ext.\ Star) \qquad (1)$$

Figure 2 displays the average asteroid NIR spectrum for each asteroid in this work, normalized, and plotted at very similar scales. Subsequent spectral analysis involves isolating the relevant spectral band parameters, such as the Band I (~0.9-μm) and Band II (~1.9-μm) pyroxene absorption band centers and areas, and the Band Area Ratio (BAR = Band II/Band I). Deriving these parameters requires isolating the individual Band I and Band II features with a linear continuum bounding each feature and terminating at the local maximum on each side of the feature (Cloutis et al., 1986). We use two independent software packages (Reddy et al., 2011a,b; Lindsay et al., 2013, 2015) to derive and compare the calculated Band I/II absorption band centers, band areas and BARs, and band depths. MATLAB analysis code from Reddy et al. (2011a,b) entails a manual process of isolating each absorption feature, applying 3[rd] and 4[th] order polynomials to the feature, repeating the process at least 10 times per absorption feature and polynomial order, and averaging the band center results. For each average Band I and Band II absorption center, error bars are determined by using the maximum individual band center deviation from the average band center. While the maximum deviation on each side of the average band center is not always equivalent, using the maximum overall deviation is a conservative approach that should best approximate the errors in band center determinations. The same process is applied to band areas and determination of the BAR for each asteroid. Table 3 displays the Band I, Band II, and BAR results from the MATLAB analysis.

The Spectral Analysis Routine for Asteroids (SARA) was also used to provide a check on the MATLAB results and to determine if both programs produce comparable results (Lindsay et al., 2013, 2015). SARA is an automated IDL program that accesses text files of asteroid spectra and conducts analyses in a manner almost equivalent to the MATLAB routines. The primary variation between the techniques used by MATLAB and SARA are determination of the Band II area. While the MATLAB Band II area determinations utilize the entire absorption feature, the SARA routines only measure the Band II feature out to 2.4-μm (Lindsay et al., 2013, 2015).

We employed the basic SARA analysis routines without applying keywords for V-type asteroids while also omitting temperature corrections, to obtain the raw results for band centers, areas, and depths. SARA applies $3^{rd}$, $4^{th}$, and $5^{th}$ order polynomials to determine band centers, but we only used the $4^{th}$ order and $5^{th}$ order polynomial results due to occasional poor fits when using $3^{rd}$ order polynomials.

Band center and BAR results using SARA and MATLAB for all of the asteroids, and their overall averages, are shown in Table 3. The results in Table 3 are temperature-corrected using the methods of Burbine et al. (2009) and Reddy et al. (2012b). Asteroid surface temperatures are estimated by the methods of Burbine et al. (2009). Inputs include the asteroid WISE albedo and heliocentric distance during the time of observation. Beaming factor and IR emissivity inputs are 1.0 and 0.9, respectively. Estimated temperatures are then input into Reddy et al. (2012b) equations to produce Band I and II corrections for eucrites and howardites, and diogenites, respectively. The initially chosen HED type is based on the position of the uncorrected band centers in the band-band plot. Variations of the Band I and Band II corrections for the HED meteorite types are up to ~0.003-microns and ~0.006 microns, respectively, for the typical temperatures experienced by the asteroids in this work (~150-190 K). Temperature corrections shift asteroid band centers to longer wavelengths and are larger for Band II centers than Band I centers. As seen in Table 3, band center and BAR values for the MATLAB and SARA routines are very similar and display consistency among the eight $V_p$-type asteroids in this study.

Band center and BAR data for each asteroid were then used to test for associations with basaltic achondrites, estimate the average surface pyroxene chemistry, and attempt assignment of each asteroid with its most likely HED meteorite analog (eucrite, howardite, or diogenite). These analyses are shown in Figure 3, Figure 4, and Figure 5, which include summary results for all of the asteroids in this work.

Figure 3 displays a Band I center vs. BAR plot, which is adapted from Cloutis et al. (1986). The different elliptical shapes correspond to the S-asteroid sub-types of Gaffey et al. (1993) while the rectangular region in the lower right portion of Figure 3 corresponds to the parameter space defined for the basaltic achondrites. Figure 4 is a Band I center vs. Band II center plot of the different HED meteorite types and the asteroids in this study. Figure 5 represents a pyroxene quadrilateral that plots the average surface pyroxene chemistries of the asteroids along with the zones for the different HED meteorite types. The totality of these spectral and mineralogical analyses, along with orbital and other physical information, is used to determine if an HED meteorite type is a probable analog to each asteroid.

4. Results.

4.1. (2011) Veteraniya.

(2011) Veteraniya has been classified as a member of the Vesta family via both dynamical and hierarchical clustering methods (Zappala et al., 1995; Mothe-Diniz et al. 2012; Nesvorny, 2012). Physical data on the asteroid include the WISE-derived diameter and albedo, which are 5.2 km and 0.463, respectively (Masiero et al., 2011). The rotational period for (2011) Veteraniya has been measured to be $8.209 \pm 0.005$ hours (Hasegawa et al., 2012). Xu et al. (1995) reported a visible-wavelength spectrum of (2011) Veteraniya that displays the short-wavelength portion of the broad ~0.9-μm pyroxene absorption feature.

The average NIR spectrum for (2011) Veteraniya is shown in Figure 2A and the observational circumstances are shown in Table 1. The Band I and II pyroxene absorption features are apparent in this high-quality average spectrum. This spectrum is an average of 10 spectra with a total integration time of 1200 seconds that were obtained on August 26, 2014 UT. The continuum-removed absorption features have Band I and II averages of 0.941- and 1.960-μm, respectively. The calculated average BAR is 2.162. Band I and II absorption band depths are ~36% and ~32%, respectively.

Plotting Band I and BAR values for (2011) Veteraniya place the asteroid in the basaltic achondrite parameter space shown in Figure 3. The pyroxene spectral calibrations from Gaffey et al. (2002) and Burbine et al. (2009) were applied to estimate the asteroid's average surface pyroxene chemistry. The overall derived average surface pyroxene chemistry for (2011) Veteraniya is $Wo_{12}Fs_{42}$ and is shown in Table 4. These data were plotted in the pyroxene quadrilateral in Figure 5 and are located in a region consistent with the howardites, but with a possible enhancement in the eucritic component due to proximity to the eucrite region in Figure 5. Therefore, the cumulative evidence currently suggests that (2011) Veteraniya is a probable Vestoid that derives from Vesta's surface and has a largely howarditic composition with a potential eucrite enhancement.

4.2. (5875) Kuga.

(5875) Kuga has not yet been categorized with an asteroid family, but exhibits orbital parameters (Table 2) that suggest a dynamical relationship with (4) Vesta. Physical data for (5875) Kuga include a rotational period of $5.551 \pm 0.002$ hours, and a WISE-derived diameter and albedo of 7.5 km and 0.381, respectively (Carbo et al., 2009; Masiero et al., 2011). No visible-wavelength or NIR spectra for (5875) Kuga have been previously reported.

The average NIR spectrum for (5875) Kuga is shown in Figure 2B and the observational circumstances are shown in Table 1. The Band I and II pyroxene absorption features are apparent in this high-quality average spectrum. This spectrum is an average of 10 spectra with a total integration time of 1200 seconds that were obtained on September 3, 2014

UT. The continuum-removed absorption features have a Band I and II average of 0.946- and 1.970-μm, respectively. The calculated average BAR is 1.918. The Band I and II absorption band depths are ~39% and ~32% respectively.

Plotting Band I and BAR values for (5875) Kuga place the asteroid in the basaltic achondrite parameter space in Figure 3. The pyroxene spectral calibrations from Gaffey et al. (2002) and Burbine et al. (2009) were used to estimate an average surface pyroxene chemistry of $Wo_{14}Fs_{44}$ (Table 4). These data plot in the eucrite region of the pyroxene quadrilateral (Figure 5). (5875) Kuga displays spectral and mineralogical properties that are most consistent with the eucrites compared to the other $V_p$-type asteroids in this work, although the data in Figure 5 does not rule out a sub-equal or lesser abundant howardite component. Therefore, the cumulative evidence currently suggests that (5875) Kuga is a probable Vestoid that derives from Vesta's surface and has a largely eucritic surface composition with a potential additional howardite component.

### 4.3. (8149) Ruff.

(8149) Ruff is another $V_p$-type asteroid that orbits in the dynamical neighborhood of (4) Vesta, but has been classified as a member of the Flora family (Nesvorny, 2012). Hardersen et al. (2014) reviewed the close proximity of the Flora and Vesta families with each other, which now makes (8149) Ruff the third $V_p$-type asteroid [along with (5235) Jean-Loup and (5560) Amytis] to be associated with the Flora family (Zappala et al., 1995). Physical data for (8149) Ruff include a WISE-derived diameter and albedo of 4.0 km and 0.582, respectively. No visible-wavelength or NIR spectra of (8149) Ruff have been previously reported.

The average NIR spectrum for (8149) Ruff is shown in Figure 2C and the observational circumstances are shown in Table 1. The Band I and II pyroxene absorption features are apparent in this high-quality average spectrum. This spectrum is an average of 10 spectra with a total integration time of 1200 seconds that were obtained on September 3, 2014 UT. The continuum-removed absorption features have Band I and II averages of 0.943- and 1.949-μm, respectively. The calculated average BAR is 2.201. Band I and II absorption band depths are ~47% and ~40%, respectively.

Plotting Band I and BAR values for (8149) Ruff place the asteroid in the basaltic achondrite parameter space in Figure 3. The average surface pyroxene chemistry is $Wo_{12}Fs_{41}$ and plots in a region of the pyroxene quadrilateral consistent with the howardites (with a possibly enhanced eucrite component) (Gaffey et al., 2002; Burbine et al., 2009) (Table 4, Figure 5). Therefore, the cumulative evidence suggests that (8149) Ruff is a probable Vestoid that derives from Vesta's surface with a largely howarditic surface composition.

### 4.4. (9147) Kourakuen.

(9147) Kourakuen is a $V_p$-type asteroid lacking a family classification with an orbit placing it in the inner regions of the asteroid belt. With a semimajor axis of 2.192 AU, (9147) Kourakuen is somewhat displaced from (4) Vesta. Physical data for (9147) Kourakuen include a WISE-derived diameter and albedo of 4.9 km and 0.242, respectively. Moskovitz et al. (2009) report a visible-wavelength spectrum of (9147) Kouraken that captures the short-wavelength portion of the Band I feature from ~0.76-μm to ~0.94-μm, which is consistent with that portion of the spectrum in Figure 2D. Solontoi et al. (2012) also reports a visible-wavelength spectrum of (9147) Kourakuen.

The average NIR spectrum for (9147) Kourakuen is shown in Figure 2D and the observational circumstances are shown in Table 1. The Band I and II pyroxene absorption features are apparent in this high-quality average spectrum. This spectrum is an average of 10 spectra with a total integration time of 1200 seconds that were obtained on August 26, 2014 UT. The continuum-removed Band I and II centers are 0.942- and 1.950-μm, respectively. The average BAR is 2.190. Band I and II band depths are ~40% and ~35%, respectively. Solontoi et al. (2012) report a Band I center of 0.915-μm. Differences between our Band I center and that from Solontoi et al. (2012) are likely due to the different methods used to measure the band center, the lack of a band center temperature correction in Solontoi et al. (2012), and the somewhat greater noise in the ~0.9-μm region of the spectrum in Solontoi et al. (2012).

Plotting Band I and BAR values for (9147) Kourakuen place the asteroid in the basaltic achondrite parameter space as shown in Figure 3. The estimated average surface pyroxene chemistry is $Wo_{12}Fs_{42}$ and is shown in Table 4 (Gaffey et al., 2002; Burbine et al., 2009). These data plot in the howardite zone of the pyroxene quadrilateral in Figure 5 and are consistent with the howardites with a possibly enhanced eucrite component.

The cumulative evidence suggests that (9147) Kourakuen may be a Vestoid that derives from Vesta's surface. The inner main-belt location for (9147) Kourakuen opens the possibility that this $V_p$-type asteroid could derive from (4) Vesta and has migrated to its current orbit due to Yarkovsky/YORP effects (Bottke et al., 2006). An intriguing, but speculative, alternative is that the asteroid could originate from another basaltic parent body distinct from (4) Vesta, but did not survive the collisional regime in the early solar system. This idea requires substantial evidence to more strongly suggest a non-Vesta origin.

### 4.5. (9553) Colas.

(9553) Colas is another $V_p$-type asteroid lacking a family classification and with an inner main belt orbit somewhat similar to that of (9147) Kourakuen. With a semimajor axis of 2.199 AU, (9553) Colas is somewhat distant from (4) Vesta. Physical data for (9553) Colas include a WISE-derived diameter and albedo of 3.8 km and 0.173, respectively. Moskovitz et al. (2009, 2011) report both a visible-wavelength and near-infrared spectrum of (9553) Colas. The visible-wavelength spectrum from Moskovitz et al. (2009) is consistent with the short-wavelength portion of the Band I feature for (9553) Colas in Figure 2E. Solontoi et al. (2012) also report a visible-wavelength spectrum for (9553) Colas.

The average NIR spectrum for (9553) Colas is shown in Figure 2E and the observational circumstances are shown in Table 1. The Band I and II pyroxene absorption features are apparent in this average spectrum that includes some point-point data scatter in the ~1.9-µm and ~2.4-µm regions due to inadequate telluric corrections. This spectrum is an average of 10 spectra with a total integration time of 1200 seconds that were obtained on September 3, 2014 UT. The average Band I and II centers are 0.931- and 1.929-µm, respectively. The average BAR is 2.652. Band I and II depths are ~44% and ~43%, respectively. Solontoi et al. (2012) report a Band I center of 0.912-µm. The discrepancy between our Band I center and that from Solontoi et al. (2012) are likely due to the different methods used to measure the band center, the lack of a band center temperature correction in Solontoi et al. (2012), and the somewhat greater noise in the ~0.9-µm region of the spectrum in Solontoi et al. (2012).

Plotting Band I and BAR values for (9553) Colas place the asteroid in the basaltic achondrite parameter space in Figure 3. The average derived surface pyroxene chemistry is $Wo_8Fs_{35}$ and is shown in Table 4. This data plots in the howardite region of the pyroxene quadrilateral adjacent to the diogenite zone in Figure 5.

Moskovitz et al. (2011) observed (9553) Colas on January 8, 2009 UT and produced an average NIR spectrum from 10 spectra with a total integration time of 2000 seconds. Their reduced NIR spectrum is very similar to that in Figure 2E. The only exception is that the NIR spectrum from Moskovitz et al. (2011) does not include the short-wavelength rollover of the ~0.9-µm feature. However, this should have a minimal impact on the derived band parameters as most of the feature is present and extends down to ~0.74-µm.  MATLAB analysis of the NIR spectrum from Moskovitz et al. (2011), identical to that described in Section 3, produce temperature-corrected Band I and II centers of 0.935- and 1.934-µm, respectively. By comparison, Moskovitz et al. (2010) report temperature-corrected band centers of 0.932-µm and 1.949-µm, respectively, along with a BAR of 2.387.

Derived band areas from the Moskovitz et al. (2011) spectrum produce a BAR value of 2.03, which is smaller than our derived BAR value, but both results are still consistent with the basaltic achondrites. The derived average surface pyroxene chemistry from the Moskovitz et al. (2011) NIR spectrum is $Wo_{10}Fs_{37}$ (Gaffey et al., 2002; Burbine et al., 2009). The band parameters used directly from Moskovitz et al. (2010) result in a pyroxene chemistry of $Wo_9Fs_{38}$. Despite the larger Band II center from Moskovitz et al. (2010), the effect on the pyroxene chemistry is minimal with only a slight increase in the Fe-content compared with our results.

Therefore, the cumulative evidence suggests that (9553) Colas may be a Vestoid that derives from Vesta's surface. Analyses of our spectrum and that from Moskovitz et al. (2010, 2011) suggest a howardite meteorite analog for (9553) Colas. There may also be a somewhat larger diogenite component within that howardite composition. As with (9147) Kourakuen, the relatively far distance of (9553) Colas to (4) Vesta could be explained by Yarkovsky/YORP drift (Bottke et al., 2006). An alternative interpretation could be that (9553) Colas originated from another basaltic parent body in the inner solar system that did not survive that early solar system formation epoch.

### 4.6. (15237) 1988 RL$_6$.

(15237) 1988 RL$_6$ is a $V_p$-type asteroid that is classified in the Vesta family (Zappala et al., 1995; Nesvorny, 2012). Physical data for (15237) 1988 RL$_6$ includes a WISE-derived diameter and albedo of 2.6 km and 0.458, respectively. No visible-wavelength or NIR spectra of (15237) 1988 RL$_6$ have been previously published.

 NIR spectra for (15237) 1988 RL$_6$ were obtained on September 3 and 4, 2014 UT. The average NIR spectrum from 9/3/2014 UT is shown in Figure 2F and the observational circumstances are shown in Table 1. For both nights, 10 spectra were obtained over a total integration time of 1200 seconds. Both average NIR spectra from each night show the dual pyroxene absorption features. Their overall average spectra are very similar with the 9/3/2014 UT average spectrum being of somewhat higher quality with less data scatter in the ~1.9-μm region.  The continuum-removed Band I and II center averages are 0.933- and 1.950-μm, respectively. The calculated average is 2.485. Band I and II absorption band depths are ~37% and ~32%, respectively. For the average 9/4/2014 UT spectrum, the average Band I and II centers are 0.930- and 1.938-μm, respectively. The average BAR is 2.308 and the Band I and II absorption feature depths are ~38% and ~33%, respectively. Band center results from both nights are consistent within the errors. The different Band II center values for the two nights are most probably attributed to the greater noise in the ~1.9-μm region of the 9/4/2014 UT average spectrum.

Plotting Band I and BAR values for (15237) 1988 RL$_6$ on both nights place the asteroid in the basaltic achondrite parameter space in Figure 3. (15237) 1988 RL$_6$ plots among the HED band-band data and trend line in Figure 4. Both band-band data points are located among the howardite meteorites. The average surface pyroxene chemistry for each night is Wo$_9$Fs$_{39}$ and Wo$_8$Fs$_{36}$, respectively, and is shown in Table 4. These data plot in a region of the pyroxene quadrilateral that is consistent with the howardites (Figure 5). Therefore, the cumulative evidence suggests that (15237) 1988 RL$_6$ is a Vestoid that derives from Vesta's surface.

### 4.7. (31414) Rotaryusa.

(31414) Rotaryusa is a $V_p$-type asteroid that has not been classified as a member of an existing asteroid family. Physical data for (31414) Rotaryusa include a WISE-derived diameter and albedo of 2.8 km and 0.222, respectively. No visible-wavelength or NIR spectra of (31414) Rotaryusa have been previously published.

NIR spectra for (31414) Rotaryusa were obtained on September 3, 2014 UT. The average NIR spectrum is shown in Figure 2G and the observational circumstances are shown in Table 1. Ten spectra were obtained over a total integration time of 1200 seconds. The average NIR spectrum shows the dual pyroxene absorption features. The continuum-removed Band I and II centers are 0.932- and 1.914-μm, respectively. The calculated average BAR is 2.401. Band I and II absorption band depths are ~45% and ~22%, respectively.

Plotting Band I and BAR values for (31414) Rotaryusa place the asteroid in the basaltic achondrite parameter space in Figure 3. (31414) Rotaryusa plots just above the diogenite region of the HED band-band data and trend line in Figure 4. The average surface pyroxene chemistry is $Wo_8Fs_{32}$, which plots in the diogenite zone of the pyroxene quadrilateral immediately adjacent to the howardite zone (Gaffey et al., 2002; Burbine et al., 2009) (Figure 5). Therefore, the cumulative evidence currently suggests that (31414) Rotaryusa is a Vestoid that derives from Vesta's surface. The diogenites, or howardites with a larger diogenite component, are the most likely meteorite analogs for (31414) Rotaryusa.

### 4.8. (32940) 1995 UW$_4$.

(32940) 1995 UW$_4$ is a $V_p$-type asteroid that has not been classified as a member of an existing asteroid family. Like (9147) Kourakuen and (9553) Colas, (32940) 1995 UW$_4$ is located further in the inner asteroid belt than most asteroids in our study with a semimajor axis of 2.189 AU (Figure 1, Table 2). Physical data for (32940) 1995 UW$_4$ include a WISE-derived diameter and albedo of 3.4 km and 0.273, respectively (Table 2). No visible-wavelength or NIR spectra of (32940) 1995 UW$_4$ have been previously published.

NIR spectra for (32940) 1995 UW$_4$ were obtained on September 4, 2014 UT. The average NIR spectrum is shown in Figure 2H and the observational circumstances are shown in Table 1. Fourteen spectra were obtained over a total integration time of 1680 seconds. The average NIR spectrum shows the dual pyroxene absorption features. The derived continuum-removed Band I and II centers are 0.934- and 1.942-μm, respectively. The calculated average BAR is 2.431. Band I and II absorption band depths are ~47% and ~46%, respectively.

Plotting Band I and BAR values for (32940) 1995 UW$_4$ places the asteroid in the basaltic achondrite parameter space as shown in Figure 3. (32940) 1995 UW$_4$ plots along, and maybe slightly above, the HED band-band data and trend line in Figure 4. The average surface pyroxene chemistry is $Wo_9Fs_{38}$ and plots in the central howardite region of the pyroxene quadrilateral (Gaffey et al., 2002; Burbine et al., 2009) (Figure 5). Therefore, the cumulative evidence suggests that (32940) 1995 UW$_4$ is probably a Vestoid that derives from Vesta's surface. The location at the boundaries of the orbital space for the $V_p$-type asteroids (Figure 1) again could simply be explained by greater Yarkovsky/YORP drift over time (Bottke et al., 2006), but also hints at a possible origin from a non-Vesta basaltic parent body.

5. Discussion and conclusions.

The eight $V_p$-type asteroids in this work display dynamical, WISE albedo, and NIR spectral characteristics (e.g., band parameters) broadly consistent with the HED meteorites. Five of the eight asteroids [(2011) Veteraniya, (5875) Kuga, (8149) Ruff, (9147) Kourakuen, (31414) Rotaryusa] plot slightly above the HED meteorite data as seen in Figure 4. The band-band positions for these asteroids suggest the possible addition of a minor, spectrally active, mafic silicate mineral phase present on these asteroids' surfaces. The most likely candidates include an olivine and/or high-Ca pyroxene phase (Gaffey et al., 2002).

The overall goal for this project is to obtain, derive, and interpret NIR spectra for ~140 $V_p$-type asteroids, of which ~90% of these asteroids are at $a < 2.5$ AU and have a potential association with (4) Vesta. The most basic objective of this effort is to determine the success rate of the $V_p$ taxonomic class in successfully identifying basaltic asteroids. Earlier work based on only a relatively small number of V-type asteroids with visible- and NIR-wavelength spectra suggested a false positive detection rate of ~9% (Hardersen et al., 2014, and references therein). Thus far in our efforts, 16 $V_p$-type asteroids have been analyzed and interpreted to have a strong, likely association with (4) Vesta. The only current uncertainty to a genetic Vesta association is for those $V_p$-type asteroids in the inner main belt that are relatively distant from (4) Vesta [(9147) Kourakuen, (9553) Colas, (32940) 1995 $UW_4$]. Their location near the inner margins of the asteroid belt could represent fragments from a completely disrupted basaltic parent body that is distinct from (4) Vesta. The anomalous basaltic achondrites provide the strongest evidence that other basaltic parent bodies populated the asteroid belt, but there are currently no remote diagnostic techniques available to distinguish basalts with similar pyroxene chemistries that originate on different parent bodies (Bland et al., 2009).

In addition, our research group is continuing to analyze NIR spectra for three $V_p$-type asteroids: (11699) 1998 $FL_{105}$, (14390) 1990 $QP_{10}$, and (67299) 2000 $GS_{95}$ that suggest a non-basaltic origin. If confirmed, then at least 16 of 19 (~84%) $V_p$-type asteroids studied in this effort thus far have been identified as having a basaltic geology and a genetic association with (4) Vesta. The predictive power of the $V_p$ taxonomy will continue to be refined as additional asteroids are analyzed in this work.

Once a reasonably large number of $V_p$-type asteroids with basaltic mineralogies and HED meteorite analogs have been identified ($N > 100$), then the newly-defined Vesta genetic family could be used to study topics such as Yarkovsky/YORP drift rates as a function of asteroid diameter and the investigation of non-Vesta basaltic parent bodies in the inner asteroid belt. For orbital locations at $a > 2.5$ AU, more secure identification of basaltic asteroids in the outer asteroid belt would better define this population of likely non-Vesta basaltic asteroids and should prompt reconsideration of early solar system heating mechanisms (Herbert et al., 1991; Grimm and McSween, 1993). [26]Al radiogenic heating models do not currently predict igneous temperatures in the outer asteroid belt (Grimm and McSween, 1993), while the T Tauri induction heating mechanism only predicts heating at such distant locations in some model-dependent scenarios (Herbert et al., 1991). Better defining the current locations of basaltic asteroids in the outer asteroid belt will allow different dynamical and thermal modeling efforts that can attempt to explain how such thermally evolved asteroids have ended up in the colder, most distant regions of the asteroid belt.

6.   Acknowledgements.

The authors gratefully thank the telescope operators, staff, and management of the NASA Infrared Telescope Facility (IRTF) for continuing to assist in supporting our research efforts. The authors also thank the reviewer for comments that improved this manuscript. PSH thanks the co-authors and Amy Mainzer, Matt Nowinski, and Gordon Gartrelle for useful comments that improved this manuscript. This work is supported by NASA Planetary Astronomy Program Grant #NNX14AJ37G. This publication makes use of data products from the Wide-Field Infrared Survey Explorer, which is a joint project of the University of California, Los Angeles, and the Jet Propulsion Laboratory/California Institute of Technology, funded by the National Aeronautics and Space Administration.

# References

Alvarez-Candal, A., Duffard, R., Lazzaro, D., Michtchenko, T.A. 2006, AA, 459, 969

Binzel, R.P, Xu, S. 1993, Science, 260, 186

Bland, P.A., Spurny, P., Towner, M.C., et al. 2009, Science, 325, 1525

Bottke Jr., W.F., Vokrouhlicky, D., Rubincam, D.P., Nesvorny, D. 2006, Ann. Rev. Earth Planet. Sci., 34, 157.

Burbine, T.H., Buchanan, P.C., Binzel, R.P., et al. 2001, MAPS, 76,761

Burbine, T.H., Buchanan, P.C., Dolkar, T., Binzel, R.P. 2009, MAPS, 44, 1331

Burns, R.G. 1993, Mineralogical Applications of Crystal Field Theory, (Cambridge, UK: Cambridge University Press)

Carbo, L., Green, D., Kragh, K., et al. 2009, MPBu, 36 152

Carvano, J., Hasselmann, P.H., Lazzaro, D., Mothe'-Diniz, T. 2010, AA, 510, 1-12, A43.

Cloutis, E.A., Gaffey, M.J., Jackowski, T.L., Reed, K.L. 1986, JGR, 91, 11641

Cochran, A.L., Vilas, F., Jarvis, K.S., Kelley, M.S. 2004, Icar, 167, 360

Consolmagno, G.J., Golabek, G.J., Turrini, D. 2015, Icar, 254, 190

Cushing, M.C., Vacca, W.D., Rayner, J.T. 2004, PASP, 116, 362

De Sanctis, M.C., Migliorini, A., Luzia Jasmin, F., et al. 2011a, AA, 533, id.A77

De Sanctis, M.C., Ammannito, E., Migliorini, A. 2011b, MNRAS, 412, 2318

De Sanctis, M.C., Ammannito, E., Capria, M.T. 2012, Science, 336, 697

Duffard, R., Roig, F. 2009, PSS, 57, 229

Duffard, R., Lazzaro, D., Licandro, J. et al. 2004, Icar, 171, 120

Duffard, R., Lazzaro, D., Licandro, J., De Sanctis, M.C., Capria, M.T. 2006, Adv. Space Res., 38, 1987

Gaffey, M.J., Burbine, T.H., Piatek, J.L., et al. 1993, Icar, 106, 573

Gaffey, M.J., Cloutis, E.A., Kelley, M.S., Reed, K.L. 2002, in Asteroids III, eds. Bottke, W.F., Jr., Cellino, A., Paolicchi, P., Binzel, R.P. (Tucson, AZ: Univ. Arizona Press), 183

Grimm, R.E., McSween, H.Y. 1993, Science, 259, 653

Hardersen, P.S., Gaffey, M.J., Cloutis, E.A., Abell, P.A., Reddy, V. 2006, Icar, 181, 94

Hardersen, P., Reddy, V., Roberts, R., Mainzer, A. 2014, Icar, 242, 269

Hasegawa, S., Miyasaka, S., Mito, H., et al. 2012, in Asteroids, Comets, Meteors 2012, Proceedings of the conference held May 16-20, 2012 in Niigata, Japan, LPI Contribution No. 1667, id.6281

Herbert, F., Sonet, C.P., Gaffey, M.J. 1991, in The Sun in Time, eds. Sonett, C.P., Giampapa, M.S., Matthews, M.S. (Tucson, AZ: Univ. Arizona Press), 710

Hicks, M.D., Buratti, B.J., Lawrence, K.J., et al. 2014, Icar, 235, 60

Kelley, M.S., Vilas, F., Gaffey, M.J., Abell, P.S. 2003, Icar, 165, 215

Klima, R.L., Pieters, C.M., Dyar, M.D. 2007, MAPS, 42, 235

Le Corre, L., Reddy, V., Nathues, A., Cloutis, E.A. 2011, Icar, 216, 376

Le Corre, L., Reddy, V., Sanchez, J.A., et al. 2015, Icar, 258, 483

Lindsay, S.S., Emery, J.P., Marchis, F., et al. 2013, AAS, DPS Meeting #45, #112.04

Lindsay, S.S., Marchis, F., Emery, J.P., Enriquez, J.E., Assafin, M. 2015, Icar, 247, 53

Mainzer, A., Masiero, J., Grav, T., et al. 2012, ApJ, 745, 7

Masiero, J.R., Mainzer, A.K., Grav, T. et al. 2011, ApJ, 741, 68

Marchi, S., McSween, H.Y., O'Brien, D.P. 2012, Science, 336, 690

Mayne, R.G., Sunshine, J.M., McSween Jr., H.Y., Bus, S.J., McCoy, T.J. 2011, Icar, 214, 147

Moskovitz, N., Jedicke, R., Willman, M. 2009, Asteroid 9553 Colas visible wavelength spectrum. EAR-A-I0039/I0576-4-SDSSMOCVSPEC-V1.0:9553_TAB, NASA Planetary Data System

Moskovitz, N.A., Willman, M., Burbine, T.H., Binzel, R.P., Bus, S.J. 2010, Icar, 208, 773

Moskovitz, N.A. 2011, Near_IR Spectrum of Asteroid 9553 Colas. EAR-A-I0046-4-IRTFSPEC-V2.0:MOSKOVITZETAL2010_9553_090108T050025_TAB, NASA Planetary Data System

Mothe'-Diniz, T., Roig, F., Carvano, J.M. 2012, Mothe-Diniz Asteroid Dynamical Families V1.1. EAR-A-VARGBDET-5-MOTHEFAM-V1.1, NASA Planetary Data System

Nathues, A., Hoffman, M., Schafer, M. 2015, Icar, 258, 467

Nesvorny, D. 2012, Nesvorny HCM Asteroid Families V2.0. EAR-A-VARGBDET-5-NESVORNYFAM-V2.0. NASA Planetary Data System

Palomba, E., Longobardo, A., De Sanctis, M.C. et al. 2015, Icar, 258, 120

Poulet, F., Ruesch, O., Langevin, Y., Hiesinger, H. 2015. Icar, 253, 364

Rayner, J.T., Toomey, D.W., Onaka, P.M., et al. 2003, PASP, 115, 362

Rayner, J.T., Onaka, P.M., Cushing, M.C., Vacca, W.D. 2004, SPIE, 5492, 1498

Reddy, V., Nathues, A., Gaffey, M.J. 2011a, Icar, 212, 175

Reddy, V., Carvano, J.M., Lazzaro, D., et al. 2011b, Icar, 216, 184

Reddy, V., Le Corre, L., O'Brien, D.P., et al., 2012a, Icar 221, 544

Reddy, V., Sanchez, J.A., Nathues, A., et al. 2012b, Icar, 217, 153

Roig, F., Nesvorny, D., Gil-Hutton, R., Lazzaro, D. 2008, Icar, 194, 125

Solontoi, M.R., Hammergren, M., Gyuk, G., Puckett, A. 2012, Icar, 220, 577

Tholen, D.J. 1984, Ph.D. thesis, University of Arizona, Tucson.

Vilas, F., Cochran, A.L., Jarvis, K.S. 2000, Icar, 147, 119

Zappala, V., Bendjoya, Ph., Cellino, A., Farinella, P., Froeschle', C. 1995 Icar, 116, 291

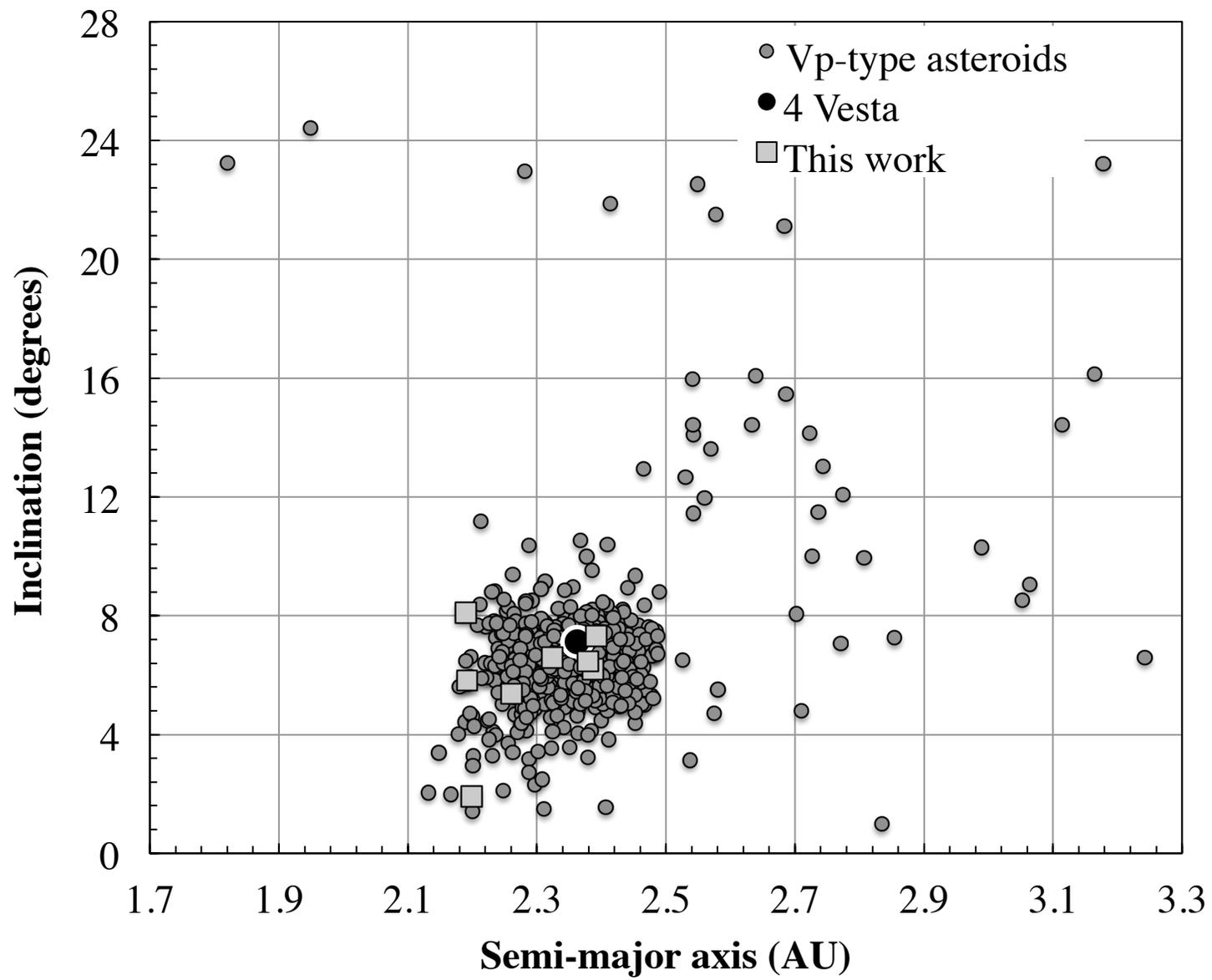

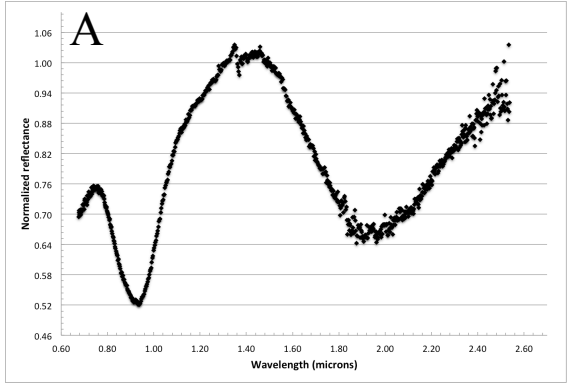
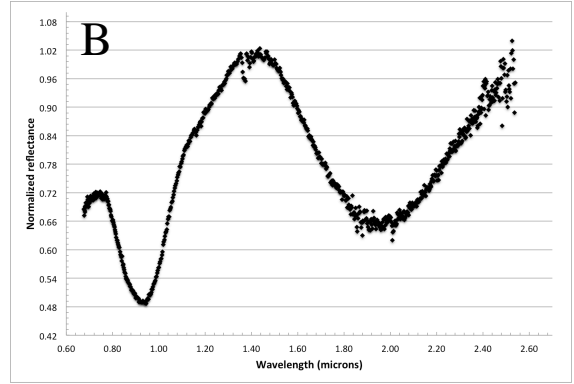
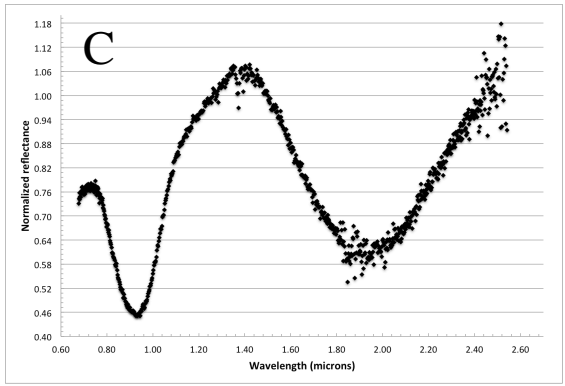
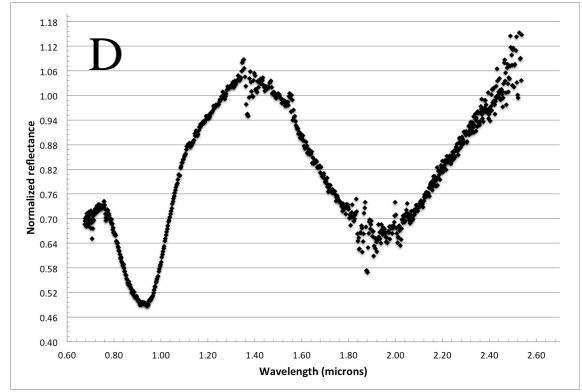
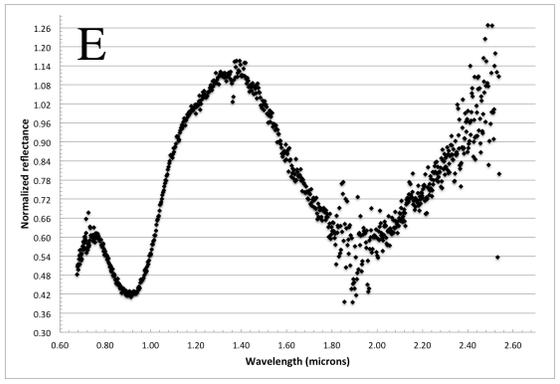
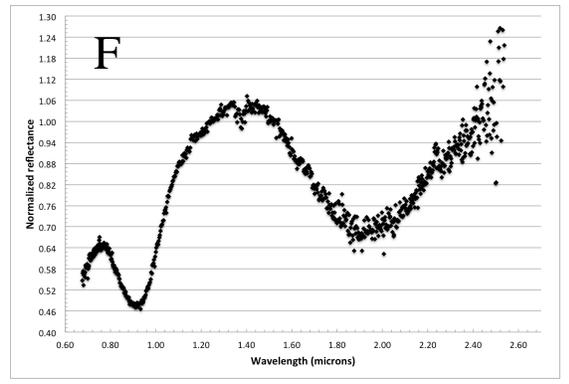
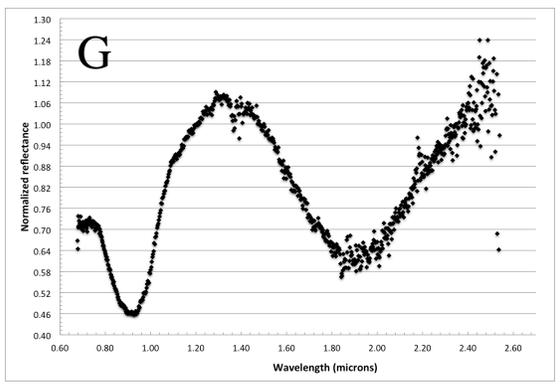
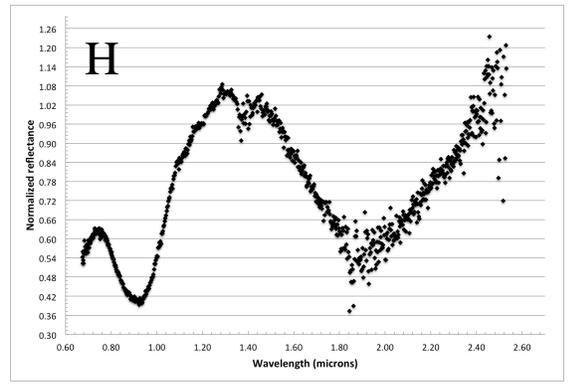

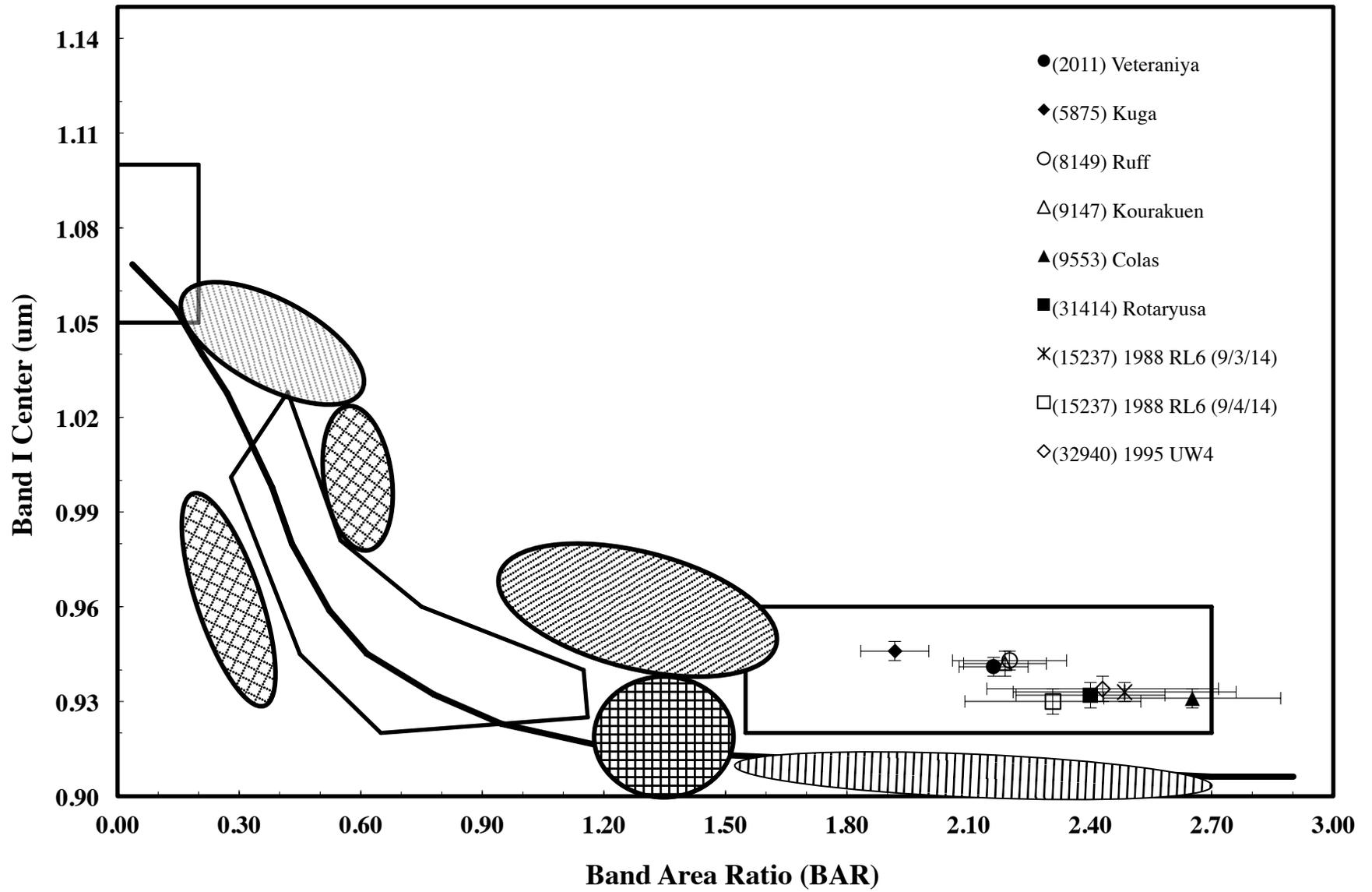

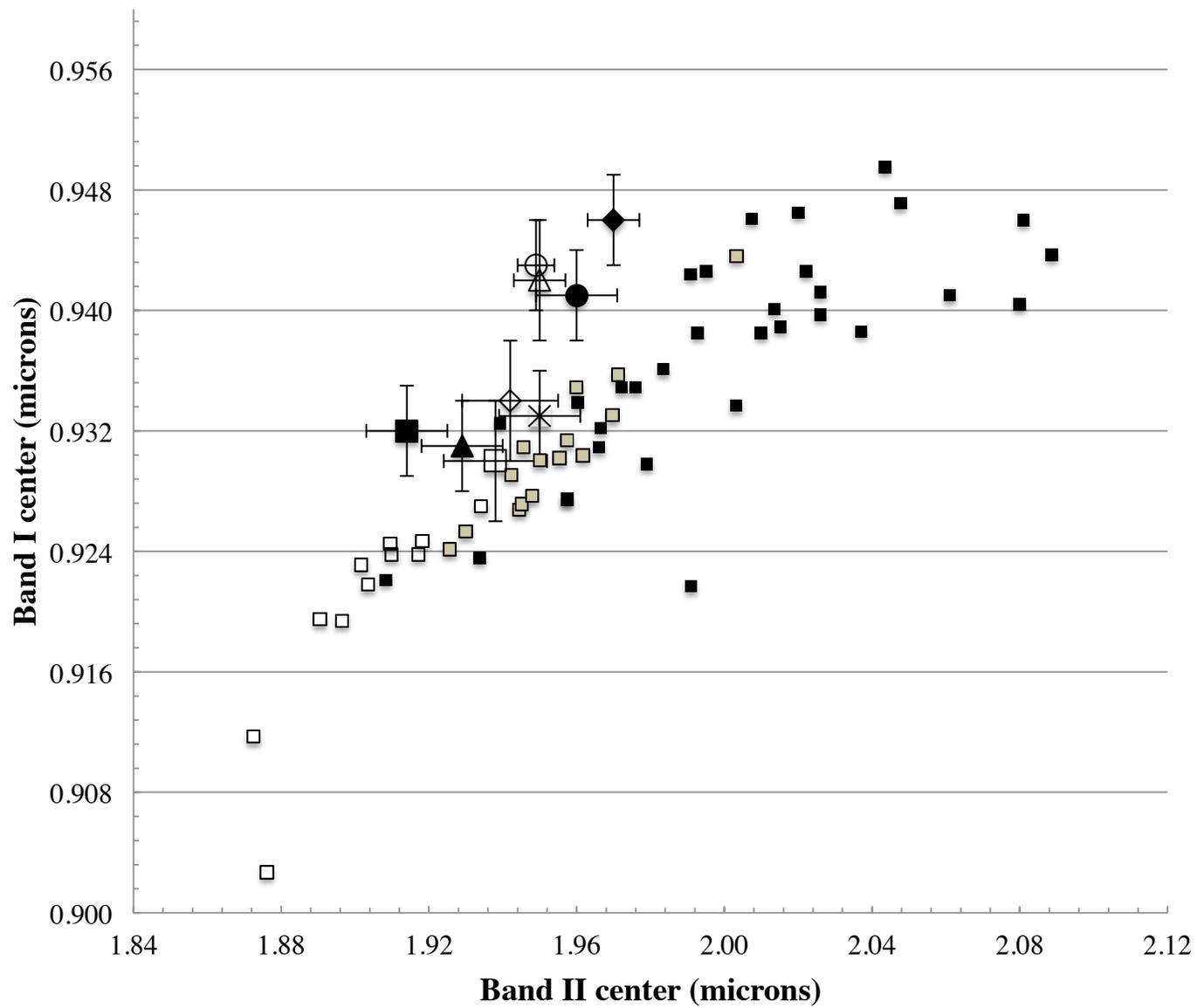

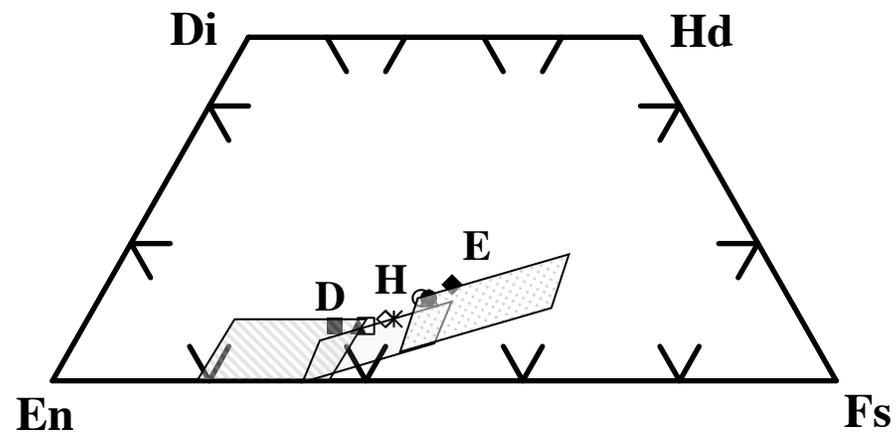

- ● (2011) Veteraniya
- ◆ (5875) Kuga
- ○ (8149) Ruff
- △ (9147) Kourakuen
- ▲ (9553) Colas
- ✳ (15237) 1988 RL6 (9/3/14)
- ▢ (15237) 1988 RL6 (9/4/14)
- ■ (31414) Rotaryusa
- ◇ (32940) 1995 UW4

| Asteroid | Observation Date (UT) | Apparent $V$ magnitude | (°) Phase angle | Total int. time (s) | Total spectra | Extinction Star | Solar Analog Star |
|---|---|---|---|---|---|---|---|
| (2011) Veteraniya | 8/26/14 | 15.12 | 8.00 | 1200 | 10 | HD 202787 (F9V) | HD 28099 (G2V) |
| (5875) Kuga | 9/3/14 | 15.47 | 3.86 | 1200 | 10 | HD 213839 (G5) | HD 28099 (G2V) |
| (8149) Ruff | 9/3/14 | 16.51 | 11.37 | 1200 | 10 | HD 4185 (G0) | HD 28099 (G2V) |
| (9147) Kourakuen | 8/26/14 | 15.59 | 3.75 | 1200 | 10 | HD 207733 (G3V) | HD 28099 (G2V) |
| (9553) Colas | 9/3/14 | 16.85 | 11.95 | 1200 | 10 | HD 203071 (G2/G3V) | HD 28099 (G2V) |
| (15237) 1988 RL6 | 9/3/14 | 16.85 | 15.12 | 1200 | 10 | HD 5172 (G0) | HD 28099 (G2V) |
| (15237) 1988 RL6 | 9/4/14 | 16.82 | 14.64 | 1200 | 10 | HD 5172 (G0) | HD 28099 (G2V) |
| (31414) Rotaryusa | 9/3/14 | 16.40 | 0.18 | 1200 | 10 | HD 216492 (G0) | HD 28099 (G2V) |
| (32940) 1995 UW4 | 9/4/14 | 16.53 | 9.84 | 1680 | 14 | HD 216432 (F8V) | HD 28099 (G2V) |

| Asteroid | *a* (AU) | *e* | *i* (°) | *WISE D$_{eff}$ (km)* | *WISE pv* | *WISE p$_{IR}$* |
|---|---|---|---|---|---|---|
| (2011) Veteraniya | 2.387 | 0.149 | 6.190 | 5.193 ± 0.384 | 0.463 ± 0.081 | -- |
| (5875) Kuga | 2.379 | 0.050 | 6.469 | 7.465 ± 0.162 | 0.381 ± 0.062 | 0.504 ± 0.115 |
| (8149) Ruff | 2.323 | 0.142 | 6.581 | 3.990 ± 0.088 | 0.582 ± 0.101 | -- |
| (9147) Kourakuen | 2.192 | 0.105 | 5.817 | 4.921 ± 0.105 | 0.242 ± 0.053 | 0.562 ± 0.069 |
| (9553) Colas | 2.199 | 0.117 | 1.920 | 3.841 ± 0.231 | 0.173 ± 0.038 | -- |
| (15237) 1988 RL6 | 2.392 | 0.149 | 7.328 | 2.590 ± 0.282 | 0.458 ± 0.099 | -- |
| (31414) Rotaryusa | 2.260 | 0.152 | 5.377 | 2.822 ± 0.094 | 0.222 ± 0.054 | -- |
| (32940) 1995 UW4 | 2.189 | 0.135 | 8.111 | 3.351 ± 0.147 | 0.273 ± 0.032 | -- |
| (4) Vesta | 2.362 | 0.090 | 7.134 | -- | -- | -- |

| Asteroid | Observation Date (UT) | MATLAB Band I (µm) | MATLAB Band II (µm) | MATLAB BAR | SARA Band I (µm) | SARA Band II (µm) | SARA BAR | Average Band I (µm) | Average Band II (µm) | Average BAR |
|---|---|---|---|---|---|---|---|---|---|---|
| (2011) Veteraniya | 8/26/14 | 0.942 ± 0.003 | 1.956 ± 0.011 | 2.235 ± 0.085 | 0.940 ± 0.003 | 1.963 ± 0.004 | 2.089 | 0.941 ± 0.003 | 1.960 ± 0.011 | 2.162 ± 0.085 |
| (5875) Kuga | 9/3/14 | 0.947 ± 0.003 | 1.968 ± 0.007 | 1.965 ± 0.084 | 0.944 ± 0.003 | 1.971 ± 0.005 | 1.870 | 0.946 ± 0.003 | 1.970 ± 0.007 | 1.918 ± 0.084 |
| (8149) Ruff | 9/3/14 | 0.942 ± 0.003 | 1.948 ± 0.005 | 2.130 ± 0.141 | 0.943 ± 0.003 | 1.950 ± 0.005 | 2.272 | 0.943 ± 0.003 | 1.949 ± 0.005 | 2.201 ± 0.141 |
| (9147) Kourakuen | 8/26/14 | 0.943 ± 0.004 | 1.947 ± 0.007 | 2.180 ± 0.102 | 0.941 ± 0.003 | 1.953 ± 0.003 | 2.200 | 0.942 ± 0.004 | 1.950 ± 0.007 | 2.190 ± 0.102 |
| (9553) Colas | 9/3/14 | 0.931 ± 0.003 | 1.926 ± 0.011 | 2.602 ± 0.218 | 0.930 ± 0.003 | 1.931 ± 0.008 | 2.702 | 0.931 ± 0.003 | 1.929 ± 0.011 | 2.652 ± 0.218 |
| (15237) 1988 RL6 | 9/3/14 | 0.933 ± 0.003 | 1.951 ± 0.011 | 2.506 ± 0.275 | 0.932 ± 0.003 | 1.949 ± 0.005 | 2.464 | 0.933 ± 0.003 | 1.950 ± 0.011 | 2.485 ± 0.275 |
| (15237) 1988 RL6 | 9/4/14 | 0.929 ± 0.004 | 1.934 ± 0.014 | 2.381 ± 0.217 | 0.930 ± 0.003 | 1.941 ± 0.008 | 2.235 | 0.930 ± 0.004 | 1.938 ± 0.014 | 2.308 ± 0.217 |
| (31414) Rotaryusa | 9/3/14 | 0.931 ± 0.003 | 1.918 ± 0.011 | 2.388 ± 0.184 | 0.932 ± 0.003 | 1.910 ± 0.005 | 2.413 | 0.932 ± 0.003 | 1.914 ± 0.011 | 2.401 ± 0.184 |
| (32940) 1995 UW4 | 9/4/14 | 0.935 ± 0.004 | 1.935 ± 0.013 | 2.460 ± 0.286 | 0.933 ± 0.003 | 1.948 ± 0.005 | 2.401 | 0.934 ± 0.004 | 1.942 ± 0.013 | 2.431 ± 0.286 |

| | MATLAB | SARA | MATLAB | SARA | |
|---|---|---|---|---|---|
| | Burbine et al. (2009) | Burbine et al. (2009) | Gaffey et al. (2002) | Gaffey et al. (2002) | |
| Asteroid | Pyx chemistry | Pyx chemistry | Pyx chemistry | Pyx chemistry | Avg. pyx chemistry |
| (2011) Veteraniya | $Wo_{10}Fs_{44}$ | $Wo_{10}Fs_{44}$ | $Wo_{14}Fs_{40}$ | $Wo_{13}Fs_{40}$ | $Wo_{12}Fs_{42}$ |
| (5875) Kuga | $Wo_{12}Fs_{48}$ | $Wo_{11}Fs_{47}$ | $Wo_{16}Fs_{40}$ | $Wo_{15}Fs_{41}$ | $Wo_{14}Fs_{44}$ |
| (8149) Ruff | $Wo_{10}Fs_{43}$ | $Wo_{10}Fs_{44}$ | $Wo_{14}Fs_{39}$ | $Wo_{14}Fs_{39}$ | $Wo_{12}Fs_{41}$ |
| (9147) Kourakuen | $Wo_{10}Fs_{44}$ | $Wo_{10}Fs_{43}$ | $Wo_{14}Fs_{39}$ | $Wo_{13}Fs_{40}$ | $Wo_{12}Fs_{42}$ |
| (9553) Colas | $Wo_{7}Fs_{36}$ | $Wo_{7}Fs_{36}$ | $Wo_{9}Fs_{33}$ | $Wo_{9}Fs_{34}$ | $Wo_{8}Fs_{35}$ |
| (15237) 1988 RL6 | $Wo_{8}Fs_{39}$ | $Wo_{8}Fs_{38}$ | $Wo_{10}Fs_{40}$ | $Wo_{10}Fs_{39}$ | $Wo_{9}Fs_{39}$ |
| (15237) 1988 RL6 | $Wo_{7}Fs_{35}$ | $Wo_{7}Fs_{37}$ | $Wo_{8}Fs_{35}$ | $Wo_{9}Fs_{37}$ | $Wo_{8}Fs_{36}$ |
| (31414) Rotaryusa | $Wo_{7}Fs_{35}$ | $Wo_{6}Fs_{34}$ | $Wo_{9}Fs_{31}$ | $Wo_{10}Fs_{29}$ | $Wo_{8}Fs_{32}$ |
| (32940) 1995 UW4 | $Wo_{8}Fs_{39}$ | $Wo_{8}Fs_{39}$ | $Wo_{11}Fs_{35}$ | $Wo_{10}Fs_{39}$ | $Wo_{9}Fs_{38}$ |

| Asteroid | HED meteorite analog |
|---|---|
| (2011) Veteraniya | Howardite (± eucrite) |
| (5875) Kuga | Eucrite |
| (8149) Ruff | Howardite (± eucrite) |
| (9147) Kourakuen | Howardite (± eucrite) |
| (9553) Colas | Howardite (± diogenite) |
| (15237) 1988 RL6 | Howardite |
| (31414) Rotaryusa | Diogenite (± howardite) |
| (32940) 1995 UW4 | Howardite |

# Table Captions

Table 1. Observational circumstances for the eight $V_p$-type asteroids in this study. Asteroid ephemeris data obtained from the JPL Horizons ephemeris service at: http://ssd.jpl.nasa.gov/?horizons. Stellar data obtained from the Simbad Astronomical Database at: http://simbad.u-strasbg.fr/simbad/.  Late F- to G-type extinction stars are observed to correct for telluric features in asteroid near-infrared (NIR) spectra. Solar analog star corrects for overall NIR spectral slope when extinctions stars are not G2V.

Table 2. Orbital and physical data for the eight $V_p$-type asteroids in this study. Orbital parameters obtained from the JPL Horizons ephemeris service at: http://ssd.jpl.nasa.gov/?horizons. Diameter and albedo information obtained from Masiero et al. (2011).

Table 3. Derived band centers and Band Area Ratio (BAR) for the eight $V_p$-type asteroids in this study. Band parameters and error bars derived from the MATLAB analysis, SARA analysis, and the average of both analyses combined (Reddy et al., 2011a,b; Lindsay et al., 2013, 2015). Overall averages used to determine average surface pyroxene chemistry for each asteroid and to test the affinities with the basaltic achondrites. Derived band parameters were broadly consistent for the MATLAB and SARA analysis techniques.

Table 4. Average surface pyroxene chemistry estimates from the eight $V_p$-type asteroids in this study. Chemistry estimates applied the mineral and basaltic achondrite meteorite calibrations from Gaffey et al. (2002) and Burbine et al. (2009), respectively. Calibrations were applied to the MATLAB and SARA band parameters data individual, and to the average of both calibrations. Results in pyroxene chemistries were broadly similar across analysis techniques and calibrations used. The errors for the derived pyroxene chemistries are $Wo_{(\pm3-4)}$ and $Fs_{(\pm4-5)}$ (Gaffey et al., (2002) and $Wo_{(\pm1.1)}$ and $Fs_{(\pm3.3)}$ (Burbine et al., 2009).

Table 5. Summary of the most likely HED meteorite analogs for the eight $V_p$-type asteroids. Meteorite analogs chosen based on the cumulative NIR spectral and mineralogical information presented herein. All eight asteroids are consistent with a basaltic achondrite interpretation. The most likely meteorite analog also chosen based on the cumulative evidence herein with howardites being the most common meteorite analog.

Figure 1. An inclination vs. semi-major axis plot for (4) Vesta, the eight $V_p$-type asteroids, and the overall 650 $V_p$-type asteroids that have been taxonomically classified (Carvano et al., 2010; Mainzer et al., 2012). The asteroids in this paper all generally display orbital parameters near Vesta, but three of the asteroids [(9147) Kourakuen, (9553) Colas, (32940) 1995 UW$_4$] appear on the inner margins ($a \sim 2.2$ AU) of the distribution of $V_p$-type asteroids.

Figure 2. Average near-infrared (NIR) reflectance spectra for each of the eight $V_p$-type asteroids. A) (2011) Veteraniya. B) (5875) Kuga. C) (8149) Ruff. D) (9147) Kourakuen. E) (9553) Colas. F) (15237) 1988 RL$_6$ (from 9/3/2014 UT). G) (31414) Rotaryusa. H) (32940) 1995 UW$_4$.

Figure 3. Band I center vs. BAR plot for the eight $V_p$-type asteroids. All eight asteroids plot within the basaltic achondrite region, which is defined by the rectangle in the lower right portion of the figure. The other regions indicate S-asteroid sub-types, ordinary chondrite, and olivine-rich asteroid regions as defined in Gaffey et al. (1993).

Figure 4. The pyroxene Band I vs. Band II plot that includes band centers for 55 HED meteorites and the eight $V_p$-type asteroids. HED meteorite data are from Le Corre et al. (2011). Band data for the asteroids plot either on or slight above the HED meteorite trend line. Positions above the trend line indicate the possible presence of minor amounts of olivine or high-Ca pyroxene, which would shift the band data above the HED trend line (Gaffey et al., 2002).

Figure 5. Pyroxene quadrilateral that defines the mineral end members for the pyroxene group minerals. The regions labeled D, H, and E represent the chemistry zones for the diogenites, howardites, and eucrites. Only one asteroid [(5875) Kuga] shows a likely eucrite analog while the remaining asteroids are largely associated with the howardites, but some with possible enhancements in the eucrite or diogenite components of their surface mineralogy. En = enstatite. Fs = ferrosilite. Di = diopside. Hd = hedenbergite.